# Cloud technologies as a tool of creating Earth Remote Sensing educational resources

Ihor V. Kholoshyn[0000-0002-9571-1758], Olga V. Bondarenko[0000-0003-2356-2674], Olena V. Hanchuk[0000-0002-3866-1133] and Iryna M. Varfolomyeyeva[0000-0002-0595-524X]

Kryvyi Rih State Pedagogical University, 54 Gagarin Ave., Kryvyi Rih, 50086, Ukraine
holoshyn@kdpu.edu.ua, bondarenko.olga@kdpu.edu.ua,
elena.ganchuk@gmail.com, iravarfolomeeva365@gmail.com

**Abstract.** This article is dedicated to the Earth Remote Sensing (ERS), which the authors believe is a great way to teach geography and allows forming an idea of the actual geographic features and phenomena. One of the major problems that now constrains the active introduction of remote sensing data in the educational process is the low availability of training aerospace pictures, which meet didactic requirements. The article analyzes the main sources of ERS as a basis for educational resources formation with aerospace images: paper, various individual sources (personal stations receiving satellite information, drones, balloons, kites and balls) and Internet sources (mainstream sites, sites of scientific-technical organizations and distributors, interactive Internet geoservices, cloud platforms of geospatial analysis). The authors point out that their geospatial analysis platforms (Google Earth Engine, Land Viewer, EOS Platform, etc.), due to their unique features, are the basis for the creation of information thematic databases of ERS. The article presents an example of such a database, covering more than 800 aerospace images and dynamic models, which are combined according to such didactic principles as high information load and clarity.

**Keywords:** Earth remote sensing data, aerospace images, ERS educational resources.

## 1 Introduction

### 1.1 Scientific relevance of the research

Nowadays, there is no doubt that the Earth Remote Sensing (ERS) data, as one of the basic directions of geoinformation technologies, is a unique source for the processes and phenomena occurring in virtually all components of the geographical envelope of the planet. In this regard, agreeing with Svetlana S. Karimova and Mikhail V. Veselov [10], Naphisa Z. Khasanshina [12; 11], we emphasize that the use of aerospace images in the educational process is not only to improve the informative content of the training, but also to contribute to the increase of students' interest in the subjects studied.

Currently, a huge collection of aerospace images has been accumulated, completely covering the entire surface of the Earth, and for many of the areas with multiple overlaping. However, despite such a large number of sources, one of the most important problems holding back the active introduction of ERS data into the educational process is the problem of providing the aerospace images with the educational process that meet the requirements for educational resources.

Ilmira B. Kiyamova [23] formulated requirements for the aerospace images selection for subject results achieving in geography training, which can be combined into three groups: scientific and pedagogical, technical and specific, due to the content of the course geography. Scientific and pedagogical requirements ensure that the content of the images corresponds to the goals and objectives of education, allow to apply appropriate methods and organizational forms of training when working with the images. Technical and specific requirements include consideration of the deciphering properties of Earth images from space, depending on the particular shooting conditions.

However, the question remains in what form, with the use of which information and communication technologies and methodological techniques it is possible to use ERS in teaching geography. Thus, the problematicity of the problem of our study is determined by the contradiction between the potential didactic capabilities of aerospace imagery, GIS data and the state of their use in the study of geography.

## 1.2 Recent research and publications analysis

Significant contributions to the development of the theory and methodology of aerospace imaging during the training process were made by Raimund Ditter, Michelle Haspel, Markus Jahn, Isabelle Kollar, Alexander Siegmund, Kathrin Viehrid, Daniel Volz [6] and Simone Naumann [20].

In the countries of post-Soviet area, the theory of geoinformation technologies was developed by Aleksandr D. Ivannikov, Vladimir P. Kulagin, Aleksandr N. Tikhonov, and Viktor Ia. Tsvetkov [9], Natalia V. Konovalova and Evgenii G. Kapralov [16], Oleksandr O. Svitlychnyi and Serhiy V. Plotnytskyi [22]. In the scientific literature the issues of geoinformation education in terms of higher education (Liliia E. Gutorova [8]) are more often considered in the context of training different future specialists, for example: GIS design and modelling (Aleksandr M. Berliant [2], Irina K. Lure and Vladimir S. Tikunov [17]); mining engineers (Vladimir S. Morkun [19], Serhiy O. Semerikov, Svitlana M. Hryshchenko, Kateryna I. Slovak [18]); future teachers (Olga V. Bondarenko, Olena V. Pakhomova, Vladimir I. Zaselskiy [4], Włodzimierz Lewoniewski [3], Ihor V. Kholoshyn, Olena V. Hanchuk and Ekateryna O. Shmeltser [13]).

Unfortunately, the use of ERS data in domestic school practice is often ignored by methodologists and practitioners. Despite the fact that Ukraine belongs to the elite cosmic powers, space technologies still cannot find a decent display in school programs. The ERS data is considered mainly as illustrative tools in presenting some topics of school geography courses. The development of this area contributes especially great work done by the Institute of progressive technologies led by Oleksandr V. Barladin [1], which found expression in a series of space atlases of different regions of Ukraine and methodological works of Liudmyla M. Datsenko and Vitalii I. Ostroukh [5]. Significantly different from other scientific works are those of Ihor V. Kholoshyn [15; 14], who focuses not only on the general information about the ERS, but also on their applied meaning and pedagogical technologies for the implementation of ERS data into the practice of modern school.

### 1.3    Article objective

The purpose of the article is to analyze the main sources of ERS that can be used in the study of geography in school practice.

## 2    Results and discussion

One of the main sources of ERS is a variety of paper media. With the advent of the first photographs of the Earth's surface in the 30's and 40's of the last century and up to now, a huge amount of aerospace images have been accumulated in the educational and scientific literature. Thousands of scientific monographs and journals have published unique images of the Earth's surface with their description and characteristics. However, for a very clear reason, their use as educational resources are extremely limited.

Aerospace imagery in the educational literature and specialized training atlases are the most common sources of ERS data in school geography. Today, virtually no textbook or school atlas is complete without the publication of aerospace photographs. Definitely, such separate, fragmented images are of the most informative value and cannot be considered as teaching aids.

The main educational resources among the sources of this type can be considered specialized satellite and complex educational and scientific atlases, which have been issued or published recently in many countries of the world. The first such publication is a training atlas published in 1982 in the Soviet Union "USSR from Space" [25]. It first collected low-resolution satellite imagery demonstrating the potential of using space technology for the national economy.

Published in Russia in 2007 "The latest world atlas with space images" is of a particular interest [24]. It consists of two parts. The first part contains maps of regions of the world, made in a scale of 1:30 000 000 and supplemented with space images of the same scale and projection. The space picture gives a visual representation of the map; the map explains the space picture. The second part of the atlas shows all the continents on a scale of 1:4 500 000. Particularly interesting cities and localities are marked on the map and presented on the following pages as detailed space images. Optically accurate meter resolution information, reproduced in GeoEye's snapshots down to the smallest detail, allows you to see and explore the nature and its landscapes, giving an idea of the major capitals of the world. And most importantly, all the pictures are provided with the text descriptions containing interesting facts and details.

Over the past twenty years, a large number of atlases containing space images have been issued in Europe and the United States. One example is the large-format atlas "One Planet, Many People: Atlas of Our Changing Environment", developed in collaboration with the US Geological Survey, NASA and the University of Maryland [21]. The atlas, which uses satellite imagery and other advanced remote sensing technologies, is designed to document global environmental changes as a result of human natural processes and activities. Most of the atlas images are taken by LANDSAT satellites.

A major obstacle to the widespread use of educational resources in a modern-day educational process is their high cost and extremely small print runs.

Apparently, aerospace images can also be obtained from a variety of individual sources. The most up-to-date and progressive approach is to recognize the possibility of obtaining space images using satellite reception stations (for example, KosmEK). They are designed to receive images of the Earth in the visible and infrared ranges transmitted from polar orbiting satellites of the NOAA, Meteor, Resource, Ocean and Sich. Up to 30 sessions can be performed on average per day. Visibility time is 6 to 15 minutes. The amount of information received in a single communication session, that is, as long as the satellite passes through the bridge type area of the station, can be 3–20 MB. The resulting image can cover vast areas across the globe, up to several million square kilometers. The images obtained can be either black and white or color in a 1:10 000 000 scale map projection.

The technology of obtaining and processing space images using the station, allows us to solve a number of important educational tasks: detection of types of cloud cover, altitude of cloud, forecast of precipitation, climate-forming factors; seasonal location and dynamics of cyclones and anticyclones development, excellent temperature characteristics of seas and lakes, fixation of fires etc.

A cheaper and more affordable way to get your own aerial imagery is to shoot the surface of the Earth from all kinds of light carriers: drones, kites and balloons. Using these fairly simple and not very expensive devices, students can independently get aerial photos of any area from a height of up to 1 km.

However, the most popular way of receiving remote sensing today is the Internet. In this case, all Internet sources can be divided into four groups.

The first group consists of various, often not specialized sites, which feature colorful and unique aerial images of high- and ultra high-resolution as visual information resources. Most of them were obtained as a result of photographing the Earth's surface by astronauts aboard orbiting stations and spacecraft (Fig. 1).

The main purpose of these images is to show the diversity and beauty of our Earth, as well as to draw the attention of the public to all kinds of problems facing humanity. Most of these images do not have an accurate mapping, often do not even contain a brief commentary, but nevertheless, taking into account their uniqueness, it is possible with some informative additions, to recommend them as a visual pictorial tool.

The second group is the sites of scientific and technical organizations and distributors, where you can view survey images, select directory images, order them, or immediately get online. Importantly, this allows you to navigate the dynamic remote sensing data market by familiarizing yourself with the characteristics of satellites, filming equipment and the product itself.

Table 1 provides examples of sites that provide free space images on the Internet. They are frequently updated, which enables us to carry out an operational monitoring of the Earth (Fig. 2).

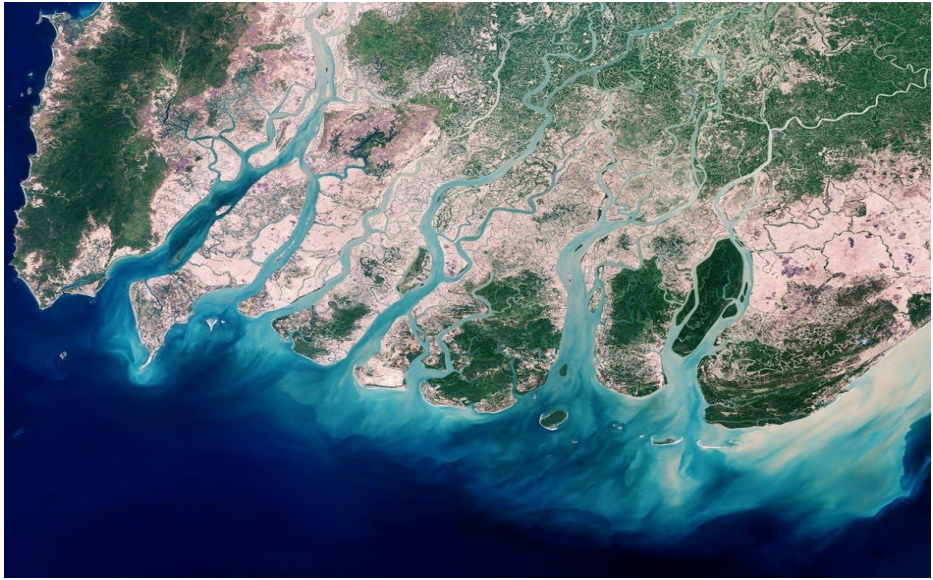

**Fig. 1.** The Iravadi River Delta (Myanmar). Satellite image, ISS

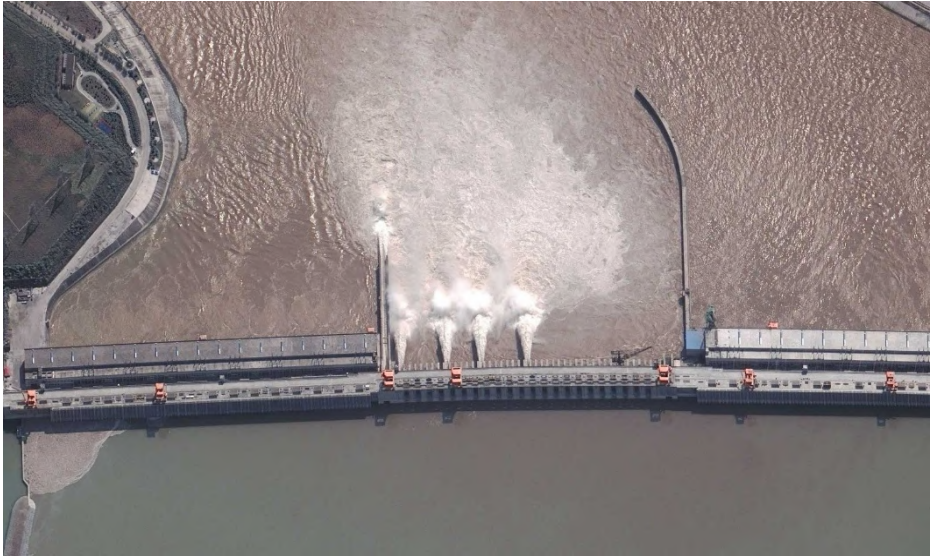

**Fig. 2.** Three Gorges Dam, China QuickBird Image: Collected September 23, 2007

The third group is represented by various interactive Internet geoservices: Google Maps, Google Earth, NASA World Wind, EarthNavigator, EarthBrowser and others. These resources allow one to get up to ten years old medium- and high-resolution images to be scaled, with the ability to scale (Fig. 3). They all offer user-friendly object search technology, as well as various additional services (related to video and photo maps, street views, directions, etc.).

**Table 1.** Examples of the sites providing free space photography on the Internet

| URL | Brief description of the image |
|---|---|
| https://www.star.nesdis.noaa.gov/GOES | Photos from the **GOES** geostationary satellite. Global coverage at very low spatial resolution (more than 1 km). Data updates every 15 minutes allowing you to track the dynamics of the atmospheric processes |
| https://eol.jsc.nasa.gov | **NASA** space images by sections: cities, natural landscapes, anthropogenic landscapes, atmospheric processes, countries of the world. Each photo is provided with geographical commentary |
| https://www.noaa.gov/satellites | **NOAA** images from polar-orbit satellites. The archive allows you to enter temporal and spatial criteria, type of equipment. Areas of application: meteorology, ecology, agriculture and forestry |
| https://glovis.usgs.gov | Relatively high-resolution space images from **LANDSAT 4-5, 7-8.** They are used for monitoring of territories and forecasting of natural anthropogenic processes |
| https://gptl.ru | Pictures obtained from Russian and foreign medium- and high-resolution satellites. They are used to solve a wide range of practical problems |
| https://earthexplorer.usgs.gov | High spatial resolution images from the **LANDSAT** satellites are used in many fields of science and economics |

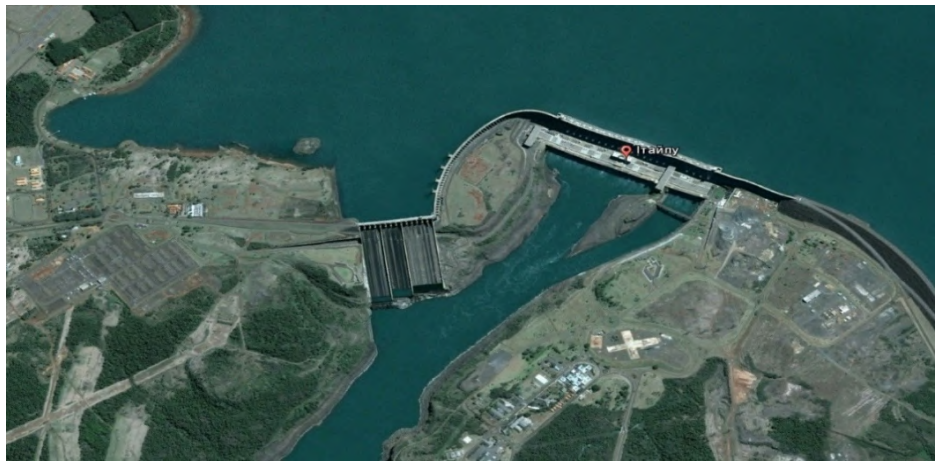

**Fig. 3.** Satellite image of the Itaipu hydroelectric station on the border of Brazil and Paraguay. Obtained using Google Earth Geosource

In particular, ignoring the details, we can say that all these geoservices have common functionality:

— visualization of the globe surface on the basis of medium-, high- and ultra high-resolution mosaic of space images (aerial images);
— easy moving around the virtual spaces of the Earth model and scaling images;
— availability of geographic tools (measuring distances, areas, determining coordinates, etc.);
— simple search services (finding different objects, routes, etc.);
— availability of custom editing tools (creating labels, lines and polygons, posting photos and comments, etc.);
— work with layers of information.

The uniqueness of such geoservices is in their interactivity, as many of them allow the user through the Internet to be not only consumers of information, but also its creators.

Like many other areas of the Internet, remote sensing also begins with the era of UGC (user-generated content) and its special case, VGI (volunteered geographic information). The emergence of wide-ranging space imagery has already led to the launch of projects that use this data as a substrate to which various geoinformation overlays are available and thousands of such projects are already being accounted for (for example, Wikimapia). In this regard, the decryption of images ceases to be the prerogative of experts and any student or pupil with a certain level of training can try to act as a decoder.

The fourth group is comprised of various cloud-based geospatial analysis platforms, which have only recently emerged but have gained widespread popularity among scientists and practitioners. First of all, it's Google Earth Engine, Land Viewer, EOS Platform and more.

These resources are free petabytes of high-resolution satellite images and have the following benefits:

1. The cloud platform provides high-speed processing of images (received, mostly, from Landsat 8 and Sentinel 2A satellites), the ability to analyze them without downloading to a personal computer.
2. The user-friendly interface allows you to find the pictures you need in a short period of time, based on their geographical location and time range.
3. A five-byte archive of publicly available remote sensing images makes it possible to find satellite images of the required area over a large time span.
4. Ability to store the desired images in the cloud.
5. Updating your snapshot database daily.
6. Selecting images from the database in different ranges (for example, Land Viewer has more than 20 combinations of ranges, such as NDVI, NBR, SAVI, etc.).

Undoubtedly, the function of comparing images taken in different time periods – time-lapse animation, is of a particular significance. The obtained dynamic models make it possible to show geographical processes in their development, help to identify the cause and effect relationships between processes and phenomena. In this way, atmospheric circulation, pollution of territories, anthropogenic changes of territories, etc. can be demonstrated (Fig. 4). The Google Earth Engine Geo-Resource site provides many classic examples of such dynamic models from satellite imagery: deforestation of the Amazon, drying out of the Aral Sea, growing the city of Las Vegas, reducing Alaska's ice cover, and more.

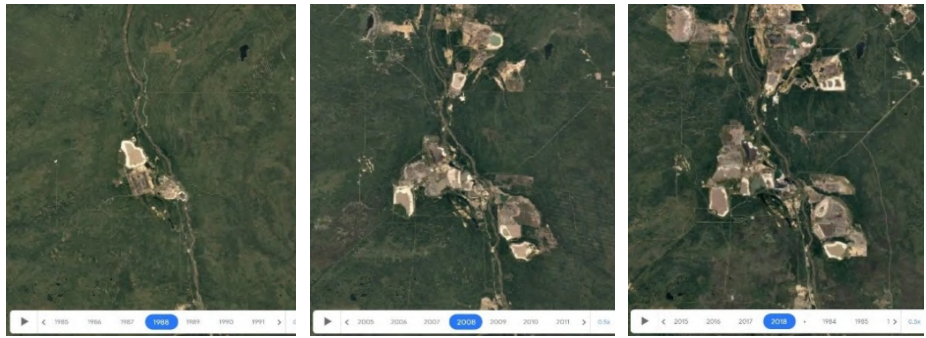

**Fig. 4.** Dynamic Model of Mine Technological Territory Change (Alberta, Canada) According to Satellite Shooting 1988, 2008, 2018 From Google Earth Engine Cloud Site

Through these examples, students are introduced to the potential of remote sensing as a tool for assessing environmental change and can be applied at different stages of the learning process, playing the role of a source of knowledge in explaining new material, as well as a means of generalizing and controlling knowledge.

However, it should be noted that the use of cloud geo-resources requires careful selection, processing and preparation, which includes their recovery and correction, transformation and decryption, and ultimately – to obtain the necessary training information. Most of our educators do not have the necessary training, and there is not enough time to carry out this painstaking work, which is a deterrent.

In this regard, teachers of higher education institutions and researchers face the issue of creating informative thematic databases of ERS with the help of which teachers can quickly and effectively use satellite imagery in the educational process.

Such systems are required to provide the following functions:

— placing pictures and their characteristics in a user-friendly format;
— systematization of data according to certain criteria;
— quick retrieval of information about snapshots stored in the database;
— search and select the information that flows upon user request;
— output information in a user-friendly form.

As an example, you can cite an educational resource in the form of a school atlas launched by ESA and Geospace [7] through Earth Observation. The atlas is built on satellite images of high spatial resolution (up to 0.6 m) and reflects the various processes that affect the development of the shells of our planet. At the same time, it should be noted that this atlas has a certain orientation and cannot be considered as an educational database of the ERS.

To solve this problem, we conducted a detailed analysis of the school curriculum geography course. As a result, possible areas of information load of the course by the data of the ERS were identified. Using a cloud platform for geospatial analysis of Google Earth Engine there has been created a collection of more than 800 aerospace images and dynamic models, combined with the principle of conformity to educational resources, namely: their high information load and visibility.

Thus, PhoA program was used to work with the collection, which is a simple and sufficiently efficient database for digital image management. Fig. 5 demonstrates the main program window. This mode is the start for the application. The main window of the program is built on the principle of standard Windows Explorer: on the left there is a group tree, and on the right it shows thumbnails and descriptions of the images in the left group.

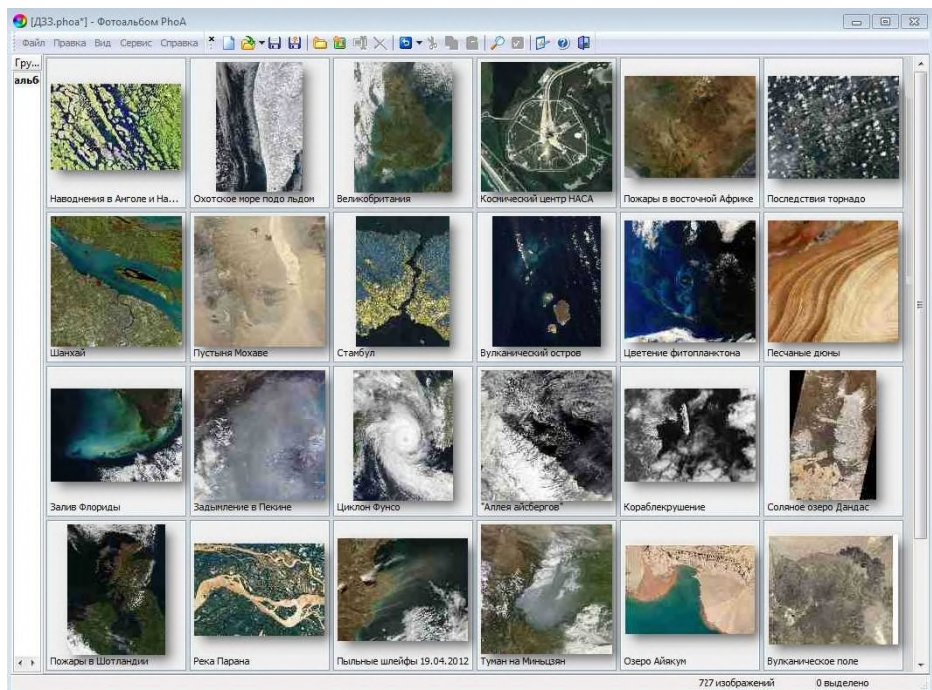

**Fig. 5.** PhoA Main Menu with aerial photos downloaded and dynamic models

The properties dialog page (Fig. 6) contains the following information: the name of the image, its geographical location, the storage folder, the source (the carrier of the recording equipment, time and shooting mode) and a brief description of the aerial photograph. The description focuses on the characterization of the depicted geographical features and their deciphering features to the extent that the picture is used as an illustration when studying individual sections or topics of school geography courses. PhoA has a fairly powerful image sampling tool. In View mode, the following functions are available: zoom (zoom in and out) of an image; scroll the image with the mouse or the keys if it is not placed completely on the screen; Go to the next, previous, first and last image in the current watch list switching from window to full screen and back; change the properties of the current image; switching the slide show on / off; displaying a description of the image; calling application settings, etc.

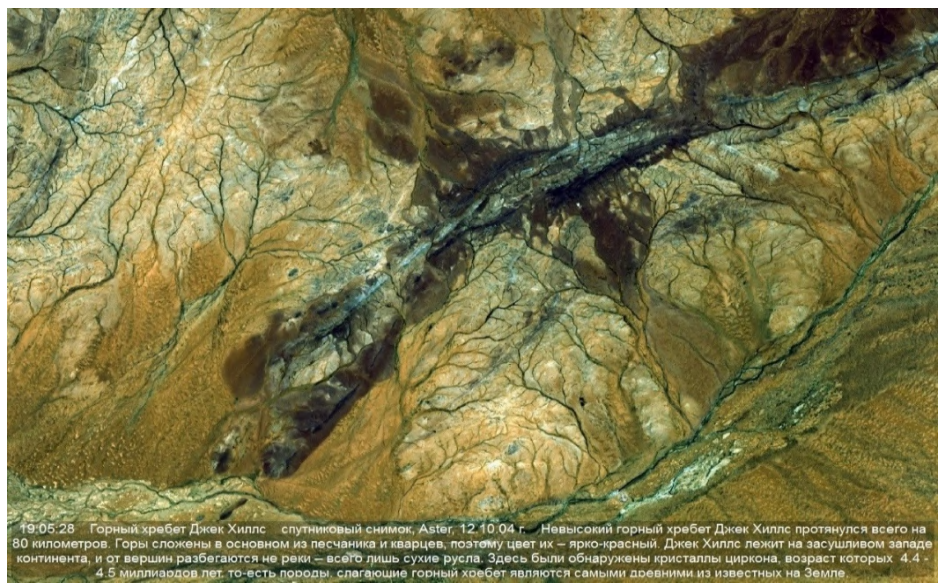

**Fig. 6.** Satellite image and its characteristics in the educational database of the ERS

However, given the huge number of images in the database (more than 800) and the inability to get information about the content of the image by its sketch, it should be recognized as more effective search, where the search criteria are the properties of the image and first of all – keywords. Keywords are a set of words that can collectively convey the semantic load of an object that they characterize. When adding snapshots to the ERS database, it is a prerequisite to enter keywords. The names of geographical objects and processes, the names of topics from sections of school geography courses, etc. are the keywords. The collection of aerial images on the cloud platform of Google Earth Engine geospatial analysis is designed primarily for teachers. The pictures and their accompanying comments are to get a highly effective visual aid when the teacher preparing for the lesson. In this case, finding the necessary pictures and analyzing them will not take much time for the teacher. The method and form of their use is similar to the application of fine arts techniques in geography classes.

## 3    Conclusions

1. ERS data are an inexhaustible source of unique information that opens doors to students into the unknown world, as their use in studying geography contributes to: a deeper understanding of the interrelations between objects and processes that take place both in society and in nature; mastering the knowledge of theoretical bases related to the introduction, storage and processing of spatial information using geoinformation technologies; formation of skills to introduce spatial information from multiple sources, organize its presentation and storage vizualize and produce results; perform the simplest operations in the analysis and synthesis of space-and-time information; use geoinformation technologies to solve a variety of daunting application problems.

2. Nowadays, the Internet remains the primary means of obtaining Earth remote sensing data. However, school teachers lack methodological guidelines for the implementation of GIS technologies when studying geography. The authors of this article tried to overcome this disadvantage by systematizing freely available Internet sources containing space images and defining their functions along with their advantages and disadvantages.

3. We see the prospects of further scientific search in covering the process of realization of practical and research orientation in geography training on the basis of data from the ERS; development of methodological course notes for practical works.